\documentclass[prb, nofootinbib, twocolumn, amsmath, amssymb, superscriptaddress]{revtex4-2}

\usepackage{bbold}
\usepackage{hyperref}
\hypersetup{colorlinks=true, citecolor=blue, linkcolor=blue, urlcolor=gray}
\usepackage{graphicx}
\usepackage{bm}
\usepackage{color}
\usepackage{url}
\usepackage{tikz}
\usetikzlibrary{calc}
\usepackage{pgfplots}
\usepackage{amsmath,amssymb}
\usepackage{verbatim}
\usepackage{physics}
\usepackage{xcolor}
\usepackage{mwe}
\usepackage{tikz}
\usetikzlibrary{calc}
\usepackage{pgfplots}

\newcommand{\bea}{\begin{eqnarray}}
\newcommand{\eea}{\end{eqnarray}}

\newcommand{\be}{\begin{equation}}
\newcommand{\ee}{\end{equation}}

\newcommand{\beal}{\begin{align}}
\newcommand{\eeal}{\end{align}}

\newcommand{\ch}{\mathcal{H}}

\begin{document}

\title{Activation of anomalous Hall effect and orbital magnetization by domain walls in altermagnets}

\author{Sopheak Sorn}
\email{sopheak.sorn@kit.edu}
\affiliation{
Institute for Quantum Materials and Technology, Karlsruhe Institute of Technology, 
76131 Karlsruhe, Germany
}%
\affiliation{Institute of Theoretical Solid State Physics, Karlsruhe Institute of Technology, 76131 Karlsruhe, Germany
}%

\author{Yuriy Mokrousov}
\affiliation{Peter Grünberg Institut (PGI-1), Forschungszentrum Jülich, 52425 Jülich, Germany}
\affiliation{Institute of Physics, Johannes Gutenberg University Mainz, 55099 Mainz, Germany}

\pacs{}
\date{\today}

\begin{abstract}
Altermagnets are an emerging class of unconventional antiferromagnets, characterized by a N\'eel ordering that does not break the translation symmetry of the underlying lattice. Depending on the orientation of the N\'eel vector, the anomalous Hall effect (AHE) may or may not exist. In the so-called pure altermagnets, AHE is forbidden by the magnetic symmetry. Here, we demonstrate that in pure altermagnets, the domain walls can lift the symmetry constraints, thereby activating the AHE and orbital magnetization.
Taking a representative example of a rutile-lattice tight-binding minimal model in slab geometry, we use the linear response theory to demonstrate the emergence of the domain wall AHE, finding that it is closely related with the orbital magnetization, while the spin magnetization does not play a significant role.
Using Landau theory, we argue that while for a random arrangement of $\pi$ domain walls, the contributions from the individual domain walls will cancel one another, an external magnetic field will favor domain-wall arrangements with specific chirality giving rise to a net AHE signal. Using group theory, we discuss how these findings can be generalized straightforwardly to certain other classes of altermagnets. Our work reveals a crucial role of the domain walls in the understanding of the Hall transport and orbital magnetism of altermagnets. Our work generally calls for a rigorous analysis of Hall-transport data for an altermagnet, ideally in conjunction with an imaging data, in order to unambiguously assign an observed Hall effect to the magnetic domains or to the domain walls.
\end{abstract}

\maketitle

\section{Introduction}
Recent developments in the field of magnetism have uncovered a new class of unconventional antiferromagnets, known as the altermagnets, wherein a N\'eel ordering together with the unusual crystal symmetry play a crucial role in dictating many physical properties~\cite{Libor2022, Libor2022_2, Bai2024}. In an altermagnet, sublattices in each unit cell are divided into two sets based on the direction of the staggering local dipole moments that define the N\'eel vector. A defining feature that distinguishes an altermagnet from a conventional antiferromagnet is the way that the two sets of the sublattices can be interchanged, neither by translation nor inversion, but by other symmetry operations such as a rotation. This prevents Kramers' degeneracy in the electronic band structures and results in a momentum-dependent band splitting, which finds applications in spintronics and drives intense research activities~\cite{Libor2022, Libor2022_2, Bai2024, Noda2016, Ahn2019, Hayami2019, Naka2019, Hayami2020}.

Due to spin-orbit coupling (SOC), the N\'eel vector is commonly locked into an easy-axis direction. In the so-called pure altermagnets, the easy axis lies in a high-symmetry direction such that the anomalous Hall effect (AHE) is forbidden by symmetry \cite{Fernandes2024}. Since a net magnetization shares the same symmetry properties as AHE, a magnetic dipole-like order parameter can be ruled out. Restricting the consideration to centrosymmetric systems, the inversion symmetry also prevents order parameters that behave like magnetic quadrupole moments~\cite{Spaldin2007, Spaldin2008, Kusunose2009, Santini2009}.
The order parameters turn out to be higher-ranked magnetic multipole moments such as magnetic octupoles and magnetic hexadecapoles \cite{Fernandes2024, McClarty2024, Spaldin2024, Steward2023}. Notable implications of the magnetic multipolar nature of the order parameters include the rise of nodal-line structures in the electronic bands \cite{Fernandes2024, Fernandes2025, Sorn2025, Zhou2024} and characteristic nonlinear response properties~\cite{Sorn2024, Cano2024, Peters2024, Ganesh2025}. For instance, magnetic octupolar order parameters leads to a third-order Hall response and a second-order magneto-electric effect, while lower orders of these effects are absent~\cite{Sorn2024, Peters2024, Ganesh2025}. Closely related to the magnetic octupolar order are such phenomena as piezomagnetism, strain-induced AHE, and unusual couplings to strain fields~\cite{Steward2023, McClarty2024, Kravchuk2024, Schmalian2025}. In contrast, when the easy axis lies in a lower-symmetry direction, the symmetry can allow for a nontrivial AHE as well as a nonzero macroscopic magnetic dipole moment---equivalently a net magnetization \cite{Libor2020, Rao1968, Comin2019, Feng2022, Libor2022AHE, Wang2023, Kluczyk2024}. The latter arises from the mechanism of a weak ferromagnetism, and its magnitude can be small. 
Such altermagents have been referred to as mixed altermagnets since ferromagnetic-like order parameters are present \cite{Fernandes2024}. Recently, it was argued that in some classes of altermagnets, in particular those with rutile structure, the weak magnetism is dominated by the $g$-tensor anisotropy  and respective anisotropy of orbital magnetism, with weak magnetization dominated by orbital contribution~\cite{Jo2025}.

\begin{figure}[t]
	\centering
	\includegraphics[width=0.8 \linewidth]{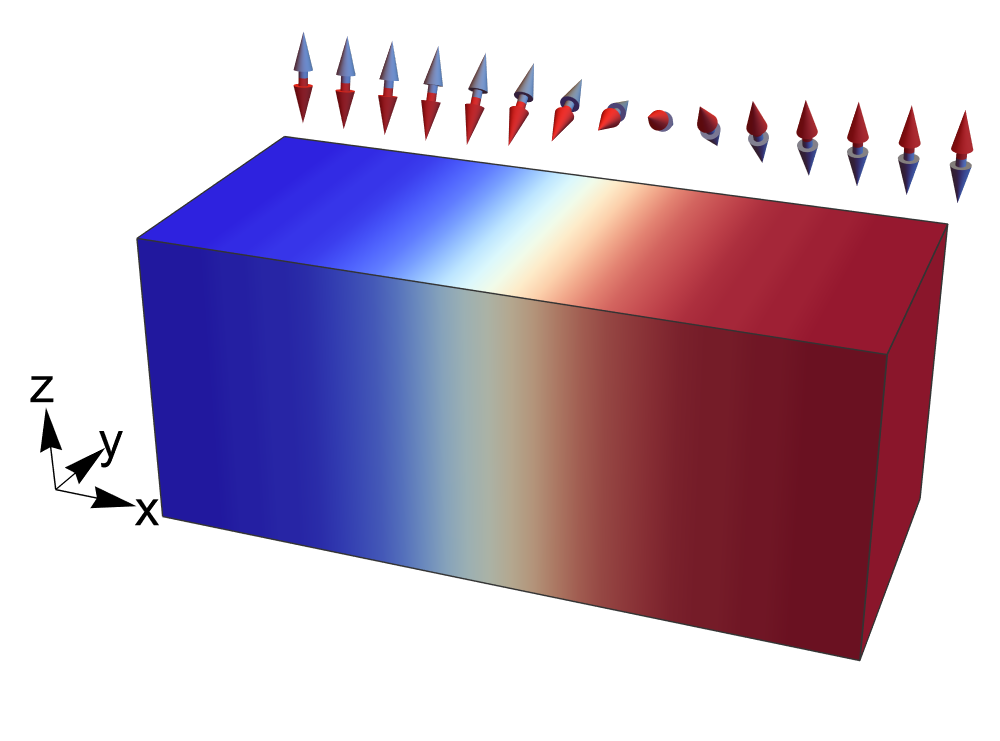}
	\caption{Geometrical setup of a film periodic in the $yz$-plane featuring an open boundary condition along the $x$-axis. The double-sided arrows denote local N\'eel vectors that form a Bloch-type domain wall along $x$.}
	\label{fig:slabgeometry}
\end{figure}

\begin{figure}[t]
    \centering
    \includegraphics[width=0.5\linewidth]{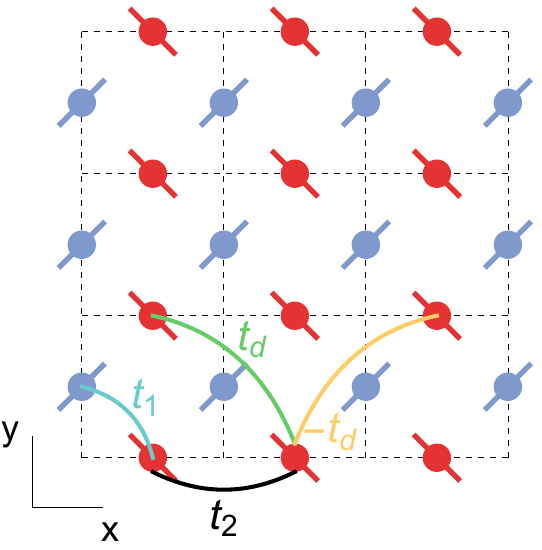}
    \caption{A representation of a rutile lattice featuring the magnetic sublattices, A (blue dots) and B (red dots). The A and the B sublattices reside on different xy-planes and experience different local potentials due to cages of non-magnetic atoms (not shown), whose orientation is shown with tilted line segments.
    }
    \label{fig:fig1}
\end{figure}

In this manuscript, we investigate Hall transport in altermagnets in the presence of magnetic domain walls (DWs), which are ubiquitous in actual settings yet have received less attention~\cite{Gomonay2024, Sorn2025}. We focus on pure altermagnets where AHE is absent in the absence of DWs, so that if AHE is activated, it is obviously attributed to the DWs.
We study this using a rutile-lattice tight-binding model on a slab geometry and coupling the electrons to a Bloch-typed DW of the N\'eel vector as schematically shown in Fig.~\ref{fig:slabgeometry}. The double-headed arrows denote the local N\'eel vectors, each of which is associated with the two antiparallel local dipole moments residing on the A and B sublattices of the rutile lattice in Fig.\ref{fig:fig1}. These local dipole moments are denoted by the blue and the red arrows. 
We assume translational invariance in the slab plane, therefore the N\'eel vector profile corresponds to an infinite DW plane. We use the Kubo formula to compute the Hall conductivity for the slab geometry and find that the DW indeed activates the AHE. AHE with a similar DW-dominating origin has been observed very recently for the first time in a layered collinear antiferromagnet EuAl$_2$Si$_2$ \cite{Xia2024}. This is in contrast to the DWs in ferromagnets, which commonly play a supplementary role since the magnetic domains can also have a nontrivial AHE contribution~\cite{Lux2020, Sorn2021, Piva2023}.

\begin{figure}[t]
	\centering
	\includegraphics[width=0.6\linewidth]{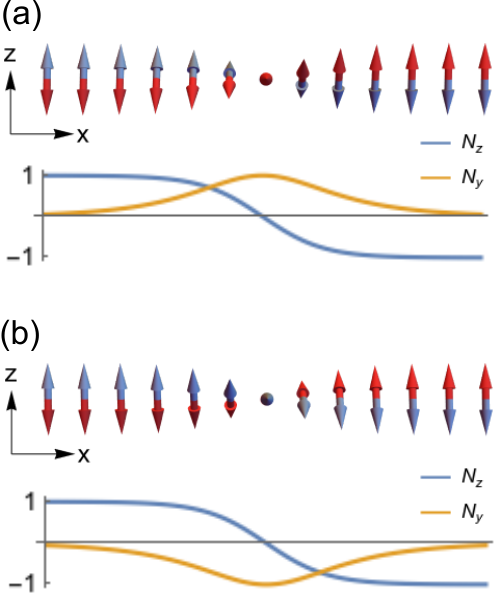}
	\caption{Illustration of Bloch-type DWs of the N\'eel vector $\vec N(\vec x)$. The orientation of the local N\'eel vector is given by $\vec N = \frac{1}{2}( \vec m_{\rm blue} - \vec m_{\rm red})$, where $\vec m_{\rm blue, red}$ denote the orientation of the local magnetic dipole moments that occupy the A and the B sublattices, respectively. Panels (a) and (b) correspond to opposite chiralities. The bottom plots show the profiles of $N_{y, z}(\vec x)$. In the presence of an inversion symmetry, the two chiralities are energetically degenerate.}
	\label{fig:dw_chirality}
\end{figure}

We find that within our model for a uniform bulk and for the easy-axis direction of $\vec N$ corresponding to a mixed altermagnet, the orbital magnetization is by far dominant over the weak spin ferromagnetism. In contrast, our analysis of the DW impact on the orbital and spin magnetization shows that they exhibit a similar order of magnitude.
Nonetheless, the computations reveal a strong connection between the orbital magnetization and the DW AHE: the Hall conductivity is most dominant in the plane perpendicular to the orbital magnetization. Meanwhile, the spin magnetization lies predominantly in the Hall plane. This means that the orientation of the spin magnetization predicts the wrong Hall plane for which the Hall effect is the strongest. Due to inversion symmetry, the DWs of opposite chirality are energetically degenerate  (see Fig.~\ref{fig:dw_chirality}). 
The AHE due to the latter has a canceling effect on that due to the former.
As a result, when we consider profiles with multiple DWs, a random arrangement of DWs with equal populations of the opposite chiralities is expected to yield a trivial AHE. However, using  Landau theory for altermagnets, we argue how an external magnetic field favors a specific chirality population, and the AHE contributions from individual DWs accumulate rather than cancel  one another. 
Meanwhile, the magnetic domains themselves are less affected due to the magnetic multipolar nature of the order parameter, which couples to the external field only at the third order, whereas the coupling with the DW chirality takes place at the linear order.
Finally, we use a group theoretical framework to generalize our results for the rutiles to other classes of pure altermagnets.

This manuscript is organized as the following. In Sec.\ref{sec:model}, we introduce the rutile-lattice tight-binding model and study the band structure, the Hall conductivity, the orbital magnetization, and the spin magnetization in the collinear bulk when DWs are absent. Section \ref{sec:dwahe} introduces a DW Ansatz in the slab geometry and discusses the Hall conductivity, the orbital magnetization, and spin magnetization induced by the DW. We also elucidate how, in multi-DW textures, AHE contributions from individual DWs may add up or cancel out one another. Section \ref{sec:Landau} discusses insights from a Landau theory for altermagnets on how an external field can select a chirality population in multi-DW textures and thus result in a nontrivial DW AHE. In Sec.\ref{sec:generalization}, we present a group-theory analysis which shows how our results for the rutile altermagnetic model can be generalized to other classes of altermagnets. Section \ref{sec:summary} is the summary.

\section{Model}
\label{sec:model}
In this Section, we introduce the tight-binding model and then analyze the band structure, the Hall conductivity, the orbital magnetization, and the spin magnetization for the collinear bulk\footnote{In the mixed altermagnetic cases, incorporating weak ferromagnetism can lead to a slight deviation from a perfect collinearity. In that case, the term ``collinear" refers to the collinearity of the sublattice exchange terms that enter the tight-binding Hamiltonian.} featuring a uniform N\'eel vector $\vec N$. These provide helpful insight for when we study the impact of DWs later.

We consider a single-orbital tight-binding model on a rutile lattice, as shown in Fig. \ref{fig:fig1}. In reciprocal space and with the basis $\left(a_{\vec k \uparrow}, a_{\vec k \downarrow}, b_{\vec k \uparrow}, b_{\vec k \downarrow}\right)^T$ of the annihilation operators, the Bloch Hamiltonian is given by \cite{Daniel2024, Fernandes2025, Sorn2025}
\bea
\label{eq:model}
\ch &=& \ch_0 + \ch_{\rm soc},\\
\ch_0 &=& - 8t_1c_{x/2}c_{y/2}c_{z/2}\tau_x - 2t'_2 c_z\tau_0 - 2t_2 (c_x + c_y)\tau_0 \nonumber\\
&& - 4 t_d s_x s_y \tau_z + J \tau_z \vec{N}\cdot \vec{\sigma}, \\
\ch_{\rm soc} &=& - 8\lambda s_{z/2} (s_{x/2}c_{y/2} \sigma_x - s_{y/2}c_{x/2} \sigma_y)\tau_y\nonumber\\
&& + 16\lambda' c_{x/2}c_{y/2}c_{z/2}(c_x-c_y)\tau_y \sigma_z,
\eea
where $c_{\alpha/n}\equiv \cos(k_{\alpha}/n)$ and $s_{\alpha/n}\equiv \sin(k_{\alpha}/n)$. $t_1$, $t_2$, $t_2'$ and $t_d$ are hopping integrals for the processes illustrated in Fig.~\ref{fig:fig1}. $\lambda$ and $\lambda'$ are the SOC-enabled hoppings. 
$\tau_i$'s and $\sigma_i$'s are Pauli matrices which act on the sublattice indices and the spin indices, respectively.
We note that $t_d$, $\lambda$ and $\lambda'$ are essential for altermagnetism since they originate in the distinct local environment of A and B sublattices.
In the latter expressions, the N\'eel vector $\vec N$ is assumed to be spatially uniform. When $\vec N$ exhibits a texture, the real-space version of the model is employed. Without loss of generality, we assume that $\vec N$ is normalized.

\begin{figure}[t]
    \centering
    \includegraphics[width=\linewidth]{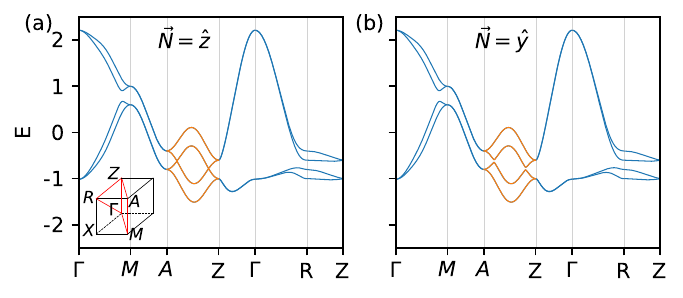}
    \caption{Bulk band structures for (a) $\vec{N}_i = \hat z$ and (b) $\vec{N}_i = \hat y$. A notable distinction between the two panels is the band anticrossings along the AZ line, which is plotted in a different color. In (a), the band crossings along AZ are protected by $z$-mirror and are a part of a Weyl nodal loop in the $k_z = \pi$ plane. In (b), the N\'eel vector lowers the symmetry, leading to the gap opening. The symmetry protection of the Weyl nodal lines such as those visible along $\Gamma$Z and $MA$, is discussed in the main text. The model parameters used are $t_1 = -0.2, t_2 = 0.025, t'_2 = -0.35, t_d = 0.15$, $J = 0.2, \lambda = - 0.025$ and $\lambda' = 0.0125,$ see main text.    }
    \label{fig:fig2}
\end{figure}

Figures \ref{fig:fig2}(a) and (b) show the band structure of the model for $\vec N = \hat z$ and $\vec N = \hat y$, respectively. As visible in (a), the band structure supports Weyl nodal lines, such as those along MA and $\Gamma$Z. On the $k_z = \pi$ plane, there are also Weyl nodal loops which intersect along AZ and manifest as two band-crossing points. The symmetry protection of these Weyl nodal structures and their physical implications, e.g. a rise of unconventional antichiral surface states, have been discussed in Ref. \cite{Fernandes2025, Sorn2025}. Particularly relevant for our AHE considerations is the Weyl nodal loops in the $k_z = \pi$ plane, which are protected by mirror $\mathcal{M}_z$ symmetry \cite{Fernandes2025}. When $\vec N = \hat y$, the mirror protection is removed, and the band-crossing points are gapped out, as can be seen in Fig.~\ref{fig:fig2}(b). These anticrossings support a large concentration of local Berry curvature which results in a nontrivial dependence of the AHE on the chemical potential, as will be seen in Sec. \ref{sec:aheuniform}. Compared with AZ, the Weyl nodal lines along $\Gamma$Z and MA seem to be intact upon the change in the direction of the N\'eel vector. A more detailed analysis reveals that, indeed, the Weyl nodal line along MA is protected by a combination of a nonsymmorphic symmetry and an antiunitary symmetry (see Appendix \ref{sec:symmetry_protection}.) However, the nodal line along $\Gamma$Z is not protected by any symmetry and can be gapped out by a symmetry-conforming perturbation (see Appendix \ref{sec:symmetry_protection}.)

\subsection{Anomalous Hall effect and orbital magnetization in
collinear bulk
}
\label{sec:aheuniform}
In this Section, we will consider the AHE in the bulk with a fixed direction of the N\'eel vector, i.e. when the DWs are absent, for two cases: (i) $\vec N = \hat z$ and (ii) for $\vec N = \hat y$. Case (i) corresponds to a pure altermagnet, where AHE is forbidden by two mirror symmetries: $S_1 = \{\mathcal{M}_x|\frac{1}{2}\frac{1}{2}\frac{1}{2}\}$ and $S_2 = \{\mathcal{M}_z|000\}$. $S_1$ requires the time-reversal-odd Hall pseudovector $\vec{\sigma}_H = (\sigma^H_{yz}, \sigma^H_{zx}, \sigma^H_{xy})^T$ to be in the $x$-direction, where $\sigma_{ij}^H$ is the Hall conductivity in the $ij-$plane. Meanwhile, $S_2$ requires  $\vec{\sigma}_H$ to be in the $z$-direction. As a result, $\vec{\sigma}_H = 0$.

\begin{figure}
	\centering
	\includegraphics[width=0.7\linewidth]{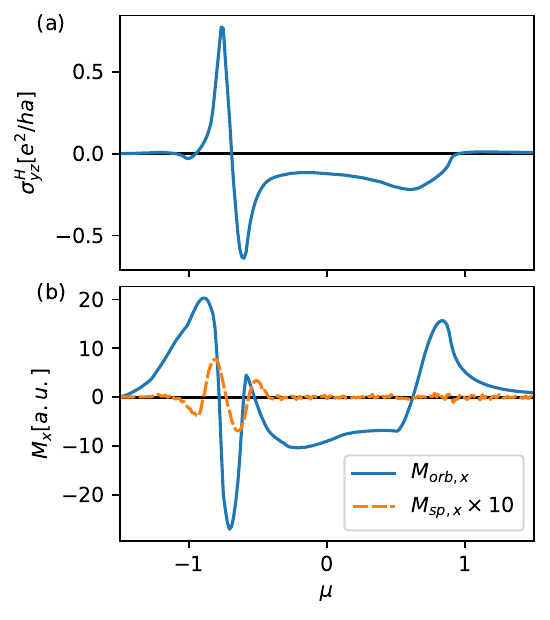}
	\caption{The chemical potential dependence of the (a) Hall conductivity $\sigma^H_{yz}$ and (b) orbital magnetization, $M_{\text{orb},x}$, and spin magnetization, $M_{\text{sp}, x}$, for collinear bulk with $\vec N = \hat y$. Here, $a$ is the length of the lattice vector along the direction perpendicular to the Hall plane. Note that the spin magnetization is multiplied with a factor of ten.}
	\label{fig:collinearbulk}	
\end{figure}

On the other hand, in case (ii), $S_2$ is broken, while $S_1$ is still present. Therefore, only $\sigma^H_{yz}$ is allowed to be nonzero. Since the average magnetization $\vec M$ is a time-reversal-odd pseudovector that transforms identically to $\vec{\sigma}_H$, its first component $M_x$ is the only allowed component. Physically, it arises in rutile structure through the orbital mechanism of $g$-factor anisotropy~\cite{Jo2025}. Case (ii), therefore, is a mixed altermagnet. In the convention of Ref.~\cite{Adamantopoulos2024}, it corresponds to a canted altermagnet, where the small canting arises from the weak ferromagnetism.

We compute the Hall conductivity $\sigma_{yz}^H$ from the $k$-dependent Berry curvature:
\bea
\sigma_{ij}^H &=& \frac{e^2}{\hbar} \sum_n \int \frac{d^3k}{(2\pi)^3} f(E_{\vec k n}) \varepsilon_{ijl}\Omega_l (\vec k n),
\eea
where $n$ is the band index, $f(E)$ is the Fermi-Dirac distribution function, $\varepsilon_{ijl}$ is the Levi-Civita symbol, and $\vec{\Omega}(\vec k n)$ is the local Berry curvature.
Figure \ref{fig:collinearbulk}(a) shows $\sigma^H_{yz}$ as a function of the chemical potential $\mu$. Near $\mu = -0.75$, $\sigma_{yz}^H$ becomes pronounced and undergoes a sign change. This behavior is closely related to the band anticrossings on the $ZA$ line in Fig.~\ref{fig:fig2}(b), which are associated with the gap opening of the Weyl nodal loops in Fig.~\ref{fig:fig2}(a). We have also checked that $\sigma^H_{xy} = \sigma^H_{zx} = 0$, consistent with the symmetry analysis.

We now compare the Hall conductivity for case (ii) with the spin magnetization $\vec M_{\rm spin}$ and the orbital magnetization $\vec M_{\text{orb}}$ \cite{Resta2005, Niu2005, Resta2006, Niu2007}:
\begin{widetext}
\bea
\label{eq:spMagColl}
M_{\text{sp}, i} &=& -\frac{e\hbar}{2} \sum_n \int \frac{d^3 k}{(2\pi)^3} f(E_{\vec k n}) \bra{u_{\vec k n}}\sigma_i \ket{u_{\vec k n}},\\
\label{eq:orbMagColl}
M_{\text{orb}, i} &=& \frac{e\hbar}{2} \sum_n \sum_{m\neq n} \int \frac{d^3 k}{(2\pi)^3}  f(E_{\vec k n}) (E_{\vec k n} + E_{\vec k m} - 2\mu) i\varepsilon_{ijl} \left[\frac{(v_j)_{\vec k}^{nm}(v_l)^{mn}_{\vec k} - (v_l)_{\vec k}^{nm}(v_j)^{mn}_{\vec k} }{(E_{\vec k n} - E_{\vec k m})^2}\right],
\eea
\end{widetext}
where we have assumed the spin $g$-factor to be 2 and the unit such that the bare electron mass $m_e$ is 1. $v_j$ is the $j$'th Cartesian component of the velocity operator, and its matrix element $(v_j)^{nm}_{\vec k} $ is given by $\bra{u_{\vec k n}}\frac{1}{\hbar} \frac{\partial \ch}{\partial k_j} \ket{u_{\vec k m} }$, where $\ket{u_{\vec k m}}$ is the Bloch eigenvector.
For the choice of the hopping integrals, the effective mass of electrons is comparable with the bare mass, which allows a direct comparison between the numerical results for $M_{\text{orb}, i}$ and $M_{\text{sp}, i}$.\footnote{The effective mass of the electrons is of the order of $\hbar^2/ a^2 W $, where $W$ denotes the bandwidth, and $a$ is the lattice constant. This means that one can vary the tight-binding parameters in order to change $W$ and the effective mass. Assuming the unit where $\hbar = a = m_e = 1$, the bandwidth $W$ as seen in Fig.\ref{fig:fig2} produces the effective mass comparable with the bare mass $m_e$. 
Therefore, we can indeed compare the order of magnitude between $\vec M_{\rm orb}$ and $\vec M_{\rm sp}$ with the choice of the tight-binding parameters.}

\begin{figure*}[t]
	\centering
	\includegraphics[width=\linewidth]{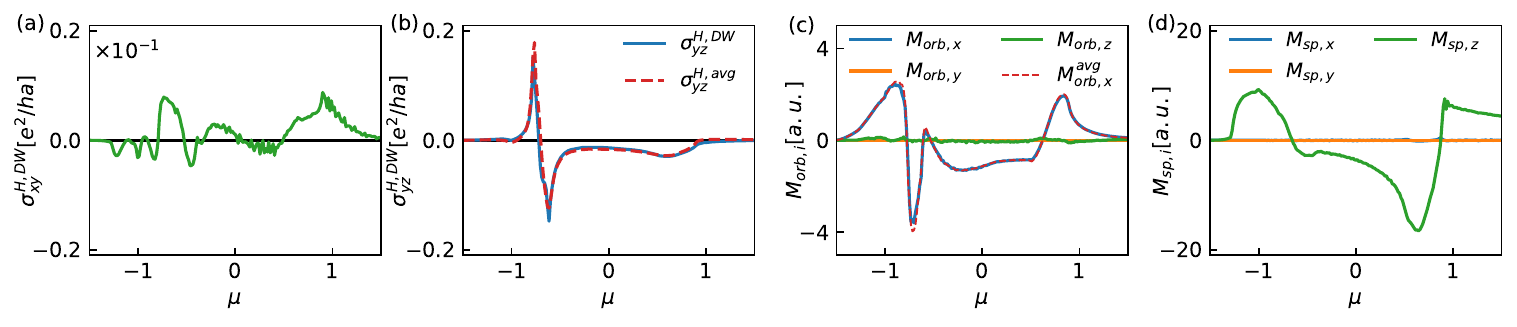}
	\caption{Anomalous Hall effect and magnetization driven by an altermagnetic domain wall. (a) $\sigma^{H, DW}_{xy}$ versus the chemical potential $\mu$. (b) $\sigma^{H, DW}_{yz}$ versus $\mu$, featuring a nontrivial dependence near $\mu = -0.75$, which is related to the anticrossings along AZ in the bulk band structure in Fig.\ref{fig:fig2}(b). (c) Orbital magnetization induced by the DW versus $\mu$. $\vec{M}_{\rm orb}$ is effectively pointing in the $x$-direction. (d) Spin magnetization induced by the DW versus $\mu$. $\vec M_{\rm sp}$ is effectively pointing in the $z$-direction. Therefore, $\vec{M}_{\rm orb}$ and $\vec{M}_{\rm sp}$ are almost perpendicular to each other. The Hall pseudovector $\vec{\sigma}_{H, DW}$, which is effectively along the $x$-axis, is closer related to the orbital magnetization. For the slab geometry, we have chosen $L_x = 100a$ and $w_{\rm dw} = 20a$. The dashed lines in (b) and (c) are obtained from a site averaging, described in the main text.}
	\label{fig:dwdata}
\end{figure*}

Figure \ref{fig:collinearbulk}(b) shows a nontrivial $\mu$-dependence of $M_{\text{sp},x}$ (dashed orange line) and $M_{\text{orb},x}$ (solid blue line), which are the only nonzero components allowed by symmetry. Note that the spin magnetization data has been multiplied by a factor of 10. Our results reveal that the orbital contribution is dominant over the spin part. 
This strongly suggests the possibility of a dominant orbital magnetism in a mixed altermagnet in general.

In the rest of the manuscript, we study how a DW in a pure altermagnet effectively realizes the scenario in the pure case (i) away from the DW region while realizing a scenario akin to the mixed case (ii) within the DW region. As a result, a nonzero AHE and concomitantly the orbital magnetization are activated and can be attributed solely to the DW.

\section{Anomalous Hall effect from altermagnetic domain walls}
\label{sec:dwahe}
To study the AHE induced by altermagnetic DWs, we consider a Bloch DW of the N\'eel vector in the slab geometry as shown in Fig.~\ref{fig:slabgeometry}. We consider the DW profile where the N\'eel vector in the $i$-th unit cell is given by $\vec N_i$, for $-L_x/2 < i < L_x/2$, and $L_x$ is the thickness of the slab. We use the following Ansatz for the N\'eel vector profile: $\vec N_i = (0, \sin \theta_i, \cos\theta_i)^T$. For the left-hand-side domain, $i\le -w_{\rm dw}/2$, $\theta_i = 0$, and for the right-hand-side domain $i \ge w_{\rm dw}/2$, $\theta_i = \pi$. In the DW region, $-w_{\rm dw}/2 < i < w_{\rm dw}/2$, $\theta_i = \pi(2i + w_{\rm dw})/2w_{\rm dw}$, where $w_{\rm dw}$ characterizes the width of the DW. We then couple the conduction electrons hopping on the rutile-lattice slab geometry with this DW profile. Reference \cite{Sorn2025} studies the electronic spectrum in the presence of such DWs, uncovering the presence of unconventional bound states at the DWs. Here, we focus on how the DW gives rise to a nontrivial anomalous Hall response.

\subsection{Anomalous Hall effect from altermagentic domain walls}
We compute the DW Hall conductivity $\sigma^{H, DW}_{ij}$ in the presence of the DW using Kubo formula and the eigenvectors of the Hamiltonian matrix for the slab-geometry problem. The full expression for the Hall conductivity can be found in the Appendix \ref{sec:Kubo} (see also Ref.~\cite{Sorn2021}). 
We will only show the results for $\sigma^{H, DW}_{yz}$ and $\sigma^{H, DW}_{xy}$, since $\sigma^{H, DW}_{zx}$ is forbidden by the symmetry operation of two-fold rotation $C_{2y}$
followed by the time reversal and a fractional translation along the $y$-direction.

Figure~\ref{fig:dwdata}(a-b) shows the chemical potential $\mu$ dependence of $\sigma^{H,DW}_{xy}$ and $\sigma^{H,DW}_{yz}$, respectively. We have used the same tight-binding parameters as in Sec.\ref{sec:model}, and we have chosen the film thickness $L_x = 100a$ and the DW width $w_{\rm dw} = 20a$, where $a$ is the length of the lattice vector.
Our general observation is that $\sigma^{H,DW}_{yz}$ has a larger magnitude compared with that of $\sigma^{H,DW}_{xy}$. The Hall pseudovector $\vec{\sigma}_{H, DW}$ is thus dominated by its first component, namely $\vec{\sigma}_{H, DW}$  effectively points in the $x$-direction. We also observe a nontrivial $\mu$ dependence of $\sigma_{yz}^{H, DW}$ which strongly resembles that in Fig.~\ref{fig:collinearbulk}(a). We compare the full result in Fig.~\ref{fig:dwdata}(b) with a site-averaged Hall conductivity defined by:
\bea
\sigma^{H, avg}_{ab} &=& \frac{1}{L_x + 1} \sum_{i = -L_x/2}^{L_x/2} \sigma^H_{ab}(\vec N_i),
\label{eq:site_avg}
\eea
where $\vec{N}_i$ is the local N\'eel vector in the $i$-th unit cell of the DW. $\sigma_{ab}^H(\vec N_i)$ is the Hall conductivity for the collinear bulk with the N\'eel vector along $\vec N_i$. The $\mu$-dependence of $\sigma_{yz}^H(\vec N_i)$'s that enter Eq.\eqref{eq:site_avg} can be found in Appendix~\ref{sec:collinearbulk}.

The result for $\sigma^{H, avg}_{yz}$ is shown as the dashed line in Fig.~\ref{fig:dwdata}(b). We find that the site averaging works remarkably well as compared to the full calculation. Based on this, one can effectively view the sign-changing behavior of $\sigma_{yz}^{H, DW}$ near $\mu=-0.75$ as the result of band anticrossings akin to the collinear-bulk cases, but here the anticrossings are caused by the DW texture. In general, we expect the site averaging to work better for a smoother DW profile since the DW-induced potential felt by the electrons varies more slowly in space. A notable consequence is that, in a smoother and correspondingly wider DW, there are more unit cells that contribute a nonzero $\sigma^H_{ab}(\vec N_i)$ in the right-hand side of Eq.~\eqref{eq:site_avg}. In other words, for a fixed $L_x$, a wider DW with a larger $w_{\rm dw}$ has nonzero contributions from more $i$-th sites in Eq.\eqref{eq:site_avg}. Consequently, the Hall conductivity is larger. Therefore, DWs with a larger width are expected to induce a larger DW AHE. 
We note also that the agreement between $\sigma^{H, avg}_{yz}$ and $\sigma^{H, DW}_{yz}$ provides strong evidence that the Hall effect is indeed activated by the DW, while bulk regions far from the DW do not contribute to the Hall effect, as expected from the altermagnetic symmetry of each domain. 
Through the definition of $\sigma^{H, avg}_{yz}$, it is also clear that an experimental setup where the Hall voltage contacts cover the whole vertical x-direction of the slab is needed in order to observe the slab-averaged DW Hall effect given by $\sigma^{H, avg}_{yz}$ and equivalently the $\sigma^{H, DW}_{yz}$.
Next, we study how the DW AHE is related to the orbital and the spin magnetization.

\subsection{Relation to the orbital and spin magnetization}
It is instructive to examine a connection between the Hall conductivity and the magnetization. Due to the slab geometry, the expressions for the orbital magnetization and the spin magnetization slightly differ from Eq.\eqref{eq:spMagColl} and \eqref{eq:orbMagColl} for the collinear bulk:
\begin{widetext}
\bea
\label{eq:spMag}
M_{\text{sp}, i} &=& -\frac{e\hbar}{2} \frac{1}{L_x + 1} \sum_n \int \frac{d^2 q}{(2\pi)^2} f(E_{\vec q n}) \bra{u_{\vec q n}}\sigma_i \ket{u_{\vec q n}},\\
\label{eq:orbMag}
M_{\text{orb}, i} &=& \frac{e\hbar}{2} \frac{1}{L_x + 1} \sum_n \sum_{m\neq n} \int \frac{d^2 q}{(2\pi)^2}  f(E_{\vec q n}) (E_{\vec q n} + E_{\vec q m} - 2\mu) i\varepsilon_{ijl} \left[\frac{(v_j)_{\vec q}^{nm}(v_l)^{mn}_{\vec q} - (v_l)_{\vec q}^{nm}(v_j)^{mn}_{\vec q} }{(E_{\vec q n} - E_{\vec q m})^2}\right],
\eea
\end{widetext}
where $\vec q = \vec k_{\parallel} = (k_y, k_z)$ is the crystal momentum in the plane of the slab, and the $j$-th Cartesian component of the velocity operator $v_j$ is defined in Appendix \ref{sec:Kubo}.

Figure \ref{fig:dwdata}(c) shows the $\mu$ dependence of $M_{\text{orb}, i}$ induced by the DW. This result shows that $\vec M_{\rm orb}$ points effectively along the $x$-axis in general. Figure~\ref{fig:dwdata}(d) illustrates the $\mu$ dependence of the spin magnetization $\vec{M}_{\rm sp}$, which is effectively pointing along the $z$-axis. With the same assumption on the effective mass as in the collinear bulk case, we can now compare the order of magnitude between $\vec{M}_{\rm orb}$ and $\vec{M}_{\rm sp}$.
Our results show that the orbital magnetization $\vec M_{\rm orb}$ is of the same order of magnitude as the spin magnetization $\vec M_{\rm sp}$. Moreover, it is clear that the orbital magnetization is a better indicator, as compared with the spin magnetization, in terms of which components of the DW Hall pseudovector $\vec \sigma_{H, DW}$ are more dominant: the dominant first component of $\vec M_{\rm orb}$ correctly indicates that the AHE is strongest in the plane perpendicular to the $x$-axis, namely $\sigma_{yz}^{H,DW}$ dominates over the other components. However, the spin magnetization, which is effectively in the $z$-direction, incorrectly suggests the dominant $\sigma_{xy}^{H, DW}$ component.

We also find that the spin magnetization exists regardless of the altermagnetism or the SOC. We have checked numerically that the spin magnetization persists even when we set $t_d = \lambda = \lambda' = 0$, which switches off the SOC and altermagnetism part in our model. In Appendix \ref{sec:nonaltermagnetic}, we explain the persistence of $\vec M_{\rm sp}$ from the gradient of the N\'eel vector and the fact that the A and B sublattices are located at slightly different positions, none of which, in contrast to  the AHE and the orbital magnetization, requires altermagnetism or SOC.
Our observation strongly indicates that the orbital magnetization is closer related to the DW AHE than the spin magnetization, in analogy to the collinear bulk case, where it is the orbital magnetization which provides the dominant contribution to the overall magnetization.

\subsection{Symmetry-imposed relations to other DW profiles}
So far, we have discussed the DW AHE for the specific DW illustrated in Fig.~\ref{fig:dw_chirality}(a). To obtain the DW AHE for the DW in Fig.~\ref{fig:dw_chirality}(b), we execute a $z$-mirror which is an operation that flips the chirality of (a) and turns it into (b). If we denote the DW Hall conductivity for the DW of type (a) by $\vec{\sigma}_{H, DW} = (\sigma_{yz}^0, \sigma_{zx}^0, \sigma_{xy}^0)^T$, then the DW Hall conductivity for type (b) is given by its $z$-mirror image
\bea
(- \sigma_{yz}^0, -\sigma_{zx}^0, \sigma_{xy}^0)^T.
\eea

There are also time-reversal counterparts of the DW profiles in Fig.~\ref{fig:dw_chirality}, which are shown in Fig.~\ref{fig:TR}. Their AHE responses are similarly obtained by implementing a time-reversal operation. Therefore, the Hall pseudovector for the DW profile in Fig.~\ref{fig:TR}(a) is given by
\bea
(- \sigma_{yz}^0, - \sigma_{zx}^0, - \sigma_{xy}^0)^T,
\eea
while that of the DW in Fig.~\ref{fig:TR}(b) is given by
\bea
(\sigma_{yz}^0, \sigma_{zx}^0, -\sigma_{xy}^0)^T.
\eea

\subsection{AHE from mutli-domain-wall profiles}
\label{sec:multiDW}
\begin{figure}
	\centering
	\includegraphics[width= 0.7\linewidth]{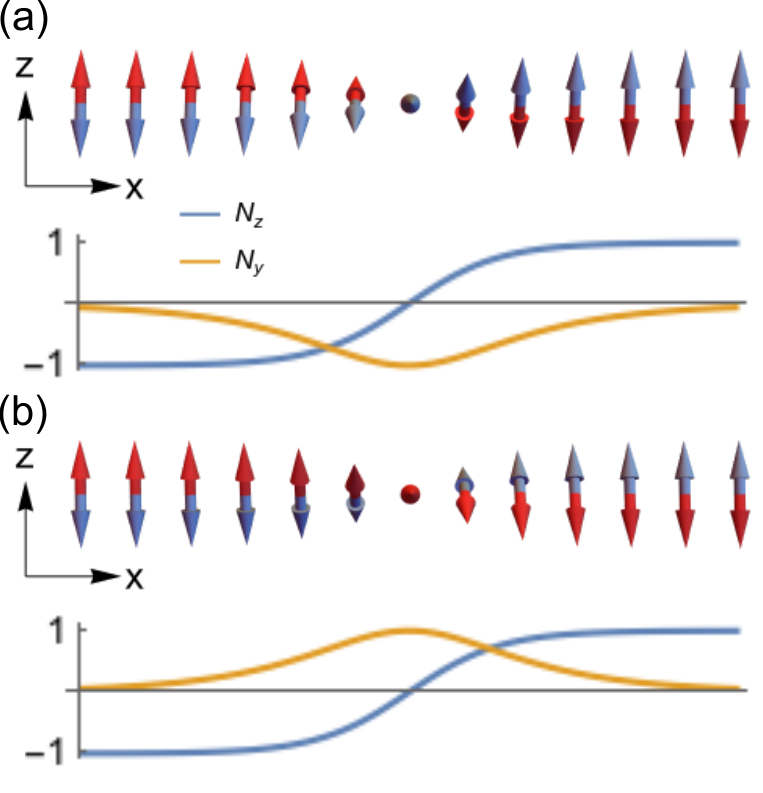}
	\caption{(a)-(b) time-reversed counterparts of Fig.~\ref{fig:dw_chirality}(a) and (b), respectively.}
	\label{fig:TR}
\end{figure}

\begin{figure}[t]
	\centering
	\includegraphics[width=0.8\linewidth]{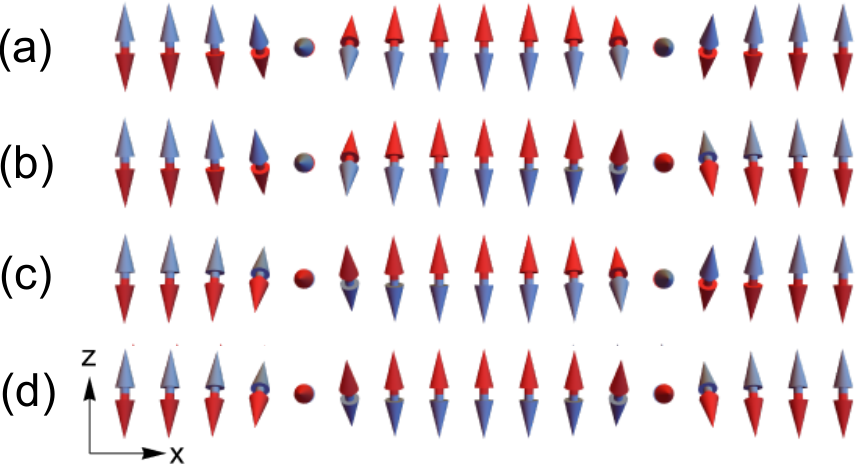}
	\caption{The four possible DW-pair combinations that are expected to arise in multiple-parallel-DW profiles.
    They differ by the combination of the sign of $N_y$ at the two DWs. (a) $N_y$ component at the two DWs has (+, +) sign. (b) (+, $-$) sign. (c) ($-$, +) sign. (d) ($-$, $-$) sign. If an applied magnetic field favors a positive $N_y$, then the combination (a) has the lowest energy.}
	\label{fig:pair}
\end{figure}

We are ready to discuss the DW AHE in the presence of multiple parallel DWs by examining how the contributions from the constituent DWs would cancel or add up with one another. This depends on which combination of the 4 DW configurations in Fig.~\ref{fig:dw_chirality} and Fig.~\ref{fig:TR} appears; see Fig.~\ref{fig:pair}. For a random combination of a large number of DWs, the contributions are expected to cancel out. As will become clear in Sec.~\ref{sec:Landau}, it may be possible to apply a magnetic field to select the sign of $N_y$ within the DW region, thereby favoring the pair in Fig.~\ref{fig:pair}(a) or Fig.~\ref{fig:pair}(d). As a result, the DW AHE contributions from the constituent DWs add up to a non-zero value. However, the DW AHE is expected to vanish for Figs.~\ref{fig:pair}(b) and (c). We also note that $\sigma_{xy}^{H, DW}$ cancels out when combining all combinations in Fig.~\ref{fig:pair}.

Finally, we estimate the order of magnitude of the DW AHE for a profile consisting of parallel DW planes as in either Fig.~\ref{fig:pair}(a) or (d) to ensure a nonzero DW AHE.
The average separation between neighboring DW planes is inversely proportional to the density of the DW, $\rho_{\rm dw}$. The net DW AHE is expected to scale like
\bea
    \sigma_{yz}^{H, DW} L_x\rho_{\rm dw},
\eea
where $\sigma_{yz}^{H, DW}$ is the Hall conductivity from the slab-geometry calculation corresponding to the slab thickness $L_x$ and a DW width $w_{\rm dw}$. 
This is expected to be a good approximation, assuming that the Hall contributions from the constituent DWs do not strongly interfere with one another, and thus these contributions simply add up. 
We note that the final outcome depends on $\rho_{\rm dw}$ and $w_{\rm dw}$. The latter is intrinsic to each DW, which determines the product $\sigma_{yz}^{H, DW} L_x$. This product is expected to converge for a sufficiently large $L_x$. In other words, we expect the AHE contribution from each DW to be mainly determined by the properties of the DW itself, which is the width $w_{\rm dw}$ in this case.
For $L_x = 100a$ and $w_{\rm dw} = 20a$, $\sigma_{yz}^{H, DW}$ is of the order of $10^{-1} e^2/ha$; see Fig.~\ref{fig:dwdata}(b). For $\rho_{\rm dw} = 1/1000\,a$, i.e. one DW per 1000 lattice constant, the total Hall conductivity is of the order of $10^{-2}e^2/h a$. For $a = 1$\,nm, the final value of the Hall conductivity is of the order of magnitude between $1$\,S/cm and $10$\,S/cm, which is an appreciable value.
Our work reveals a non-negligible role of DWs in understanding Hall transport phenomena in altermagnets, which is particularly crucial for pure altermagnets since the bulk magnetic domains do not support AHE. Therefore, our work generally calls for a proper analysis of the Hall transport data, ideally simultaneously with magnetic imaging data, in order to correctly attribute the observed Hall effect to the bulk magnetic domains or to DWs.
In the next section, we return to the demonstration of how an applied magnetic field may indeed select the DW patterns that lead to a nonzero $\sigma_{yz}^{H, DW}$. 

\section{Insights from Landau theory}
\label{sec:Landau}
In this Section, we employ the Landau theory for rutile altermagnets to explain the qualitative features of our numerical results and to outline a setup in order to achieve a nontrivial DW AHE from multi-DW configurations.

\begin{table*}[t]
    \centering
    \begin{tabular}{l l l l}
    \hline
    Point Group &  Staggering Irr. Rep. $\Gamma_S$  &  Lifshitz Invariant & $\vec N$ in domain interiors\\ \hline\hline
    $mmm$       &  $B_{1g}$            &  $\alpha_1 N_x M_y + \alpha_2 N_y M_x$	& along $\hat{z}$\\
                &  $B_{2g}$            &  $\alpha_1 N_x M_z + \alpha_2 N_z M_x$  & along $\hat{y}$ \\
                &  $B_{3g}$            &  $\alpha_1 N_y M_z + \alpha_2 N_z M_y$  & along $\hat{x}$ \\ \hline
    $4/m$       &  $B_g$               &  $\alpha_1(N_x M_x - N_y M_y)  + \alpha_2(N_xM_y + N_yM_x)$ 	& along $\hat{z}$\\ \hline
    $4/mmm$     &  $B_{1g}$            &  $\alpha_1(N_y M_y - N_x M_x)$	& along $\hat{z}$ \\
                &  $B_{2g}$            &  $\alpha_1(N_yM_x + N_xM_y)$ 	& along $\hat{z}$ \\ \hline
    \end{tabular}
    \caption{Three dimensional point groups which can support centrosymmetric pure altermagnetism exhibiting a linear coupling between $\vec N$ and $\vec M$. Other point groups may support nonlinear coupling beyond the bilinearity. The results are adapted from Ref.~\cite{Schiff2024} }.
    \label{tab:pointgroup}
\end{table*}

We consider an important aspect of the Landau theory for altermagnets formulated in Refs.~\cite{McClarty2024, Schiff2024}: coupling between N\'eel vector $\vec N$ and the magnetization $\vec M$. They are viewed as order parameters that appear in the free energy for the system. In particular, we are interested in the symmetry-allowed bilinear couplings between the components of $\vec N$ and $\vec M$. For rutile altermagnets, such a coupling appears as the following Lifshitz invariant \cite{McClarty2024, Fernandes2025}:
\bea
L_{\rm int} &=& \alpha_1 (M_x N_y + M_y N_x).
\label{eq:Lifshitz}
\eea
$\alpha_1$ is a coupling constant. We emphasize that this is the only bilinear coupling. Note also that $N_z$ and $M_z$ do not appear at the bilinear order, but they may appear in nonlinear terms. Equation \eqref{eq:Lifshitz} implies that a nonzero $M_y$ ($M_x$) can be induced by a nonzero $N_x$ ($N_y$)---the mechanism of weak ferromagnetism discussed earlier~\cite{Adamantopoulos2024,Jo2025}. 
Indeed, this explains the rise of the orbital magnetization $\vec M_{\rm orb}$ in Fig.~\ref{fig:dwdata}(c) from the nonzero $N_y$ component within the DW region \footnote{The origin of the $x$-component of $\vec M_{\rm sp}$ can also be understood this way, but the spin magnetization is dominated by the nonrelativistic contribution, i.e. the $x$-component is masked by the $z$-component which has a nonrelativistic and a non-altermagnetic origin, as discussed in the Appendix \ref{sec:nonaltermagnetic}. For this reason, the Landau theory appears less relevant for $\vec M_{\rm sp}$. }. 
It also explains why the DW gives rise to a nonzero AHE since the Hall pseudovector transforms like a magnetization vector.
We note that the Lifshitz invariant of Eq.\eqref{eq:Lifshitz} is not present in Ref.~\cite{Gomonay2024} due to the absence of SOC in the study.

Since an external magnetic field $\vec B$ also transforms like $\vec M$, one can  replace $\vec M$ in Eq.~\eqref{eq:Lifshitz} by $\vec B$. This implies that an external magnetic field directly couples to the N\'eel vector, allowing for a field control over the DW pattern. 
For example, a magnetic field in the $x$-direction couples linearly with $N_y$ and will thus energetically favor either the DW pair configuration in Fig.~\ref{fig:pair}(a) or (d). This ultimately leads to a nontrivial AHE from DWs, as claimed earlier in Sec.\ref{sec:multiDW}. We note that the magnetic domains themselves are less affected by the external magnetic field since the order parameter, i.e. $N_z$, behaves like a magnetic ocupole in the rutile lattice \cite{Spaldin2024, McClarty2024}. That is, $N_z$ transforms identically as a magnetic octupole moment under the symmetry operations of the point group and couples to the external field, beginning only at the third order: a Lifshitz invariant of the form $N_zB_xB_yB_z$ is allowed by symmetry \cite{Spaldin2024, McClarty2024}. Meanwhile, the coupling with the DW chirality already occurs at the linear order.

\section{Generalization beyond rutile altermagnets}
\label{sec:generalization}

So far, we have discussed a specific case of pure rutile altermagnets where AHE is symmetry-forbidden within the interior of each magnetic domain, and how altermagnetic DWs lift the symmetry constraint and activate the AHE. In this Section, we discuss how a similar situation can arise in other pure altermagnets. For simplicity, we restrict our attention to pure altermagnets in which it is possible to have a bilinear coupling between $\vec N$ and $\vec M$ of the following form 
\bea
L_{\rm int} = \sum_{\alpha,\beta=x,y,z} c_{\alpha\beta} N_{\alpha}M_{\beta}, 
\eea
which is a more general version of Eq.~\eqref{eq:Lifshitz} for rutiles with coupling constants  $c_{\alpha\beta}$. It is expected that the bilinearity of the coupling enables an efficient manipulation of the DW configuration by an external field $\vec B$, since nonlinear couplings could, in general, lead to a less efficient field control, e.g. requiring a large external field.

We find three crystallographic point groups in three dimensions which support pure altermagnetism and the bilinear couplings simultaneously; see Table \ref{tab:pointgroup}. This table is adapted from a recent work on Landau theories for altermagents \cite{Schiff2024}, and we will describe how to use it in the following. The first column is the crystallographic point group of the underlying lattice. To describe the physical meaning of the second column, we first recall that an altermagnetic ordering does not break the translational symmetry of the lattice. It is instructive to picture how the staggering dipole moments associated with the N\'eel order occupy and divide the sublattices in each unit cell into two groups. We define a sign-alternating pattern by assigning the value +1 to one group and $-$1 to the other group. It is essential that a translation or an inversion operation does not interchange the sublattices between the two groups to avoid the Kramers' degeneracy. Otherwise, we will get the conventional antiferromagnetism, i.e. no momentum-dependent band splitting. Nevertheless, the two groups may be interchanged by other operations, e.g. a rotation operation of the crystallographic point group \footnote{In practice, one needs to consider symmetry operations of the \emph{space} group, but each of such operations can be identified unambiguously with an operation of the point group.}. As a result, such a rotation reverses the sign of the sign-alternating pattern. While some operations of the point group change the sign of the sign-alternating pattern, other operations do not. This means that the sign-alternating pattern transforms according to a one-dimensional irreducible representation of the point group which defines $\Gamma_S$. $\Gamma_S$ varies, depending on the type of Wykoff positions taken by the sublattices. For  rutiles, the point group is $4/mmm$. Within each unit cell, the A sublattice constitutes the first group, while the B sublattice forms the second group. A four-fold rotation is an example that interchanges the two groups. The sign-alternating pattern transforms according to $\Gamma_S = B_{2g}$. Accordingly, the bilinear coupling can be read off from the third column of the table and is indeed given by Eq.~\eqref{eq:Lifshitz}. Finally, the fourth column specifies the high-symmetry direction for $\vec N$ corresponding to the pure altermagnetism.
Table \ref{tab:pointgroup} demonstrates the presence of multiple examples of pure altermagnets where the bilinear couplings are possible. Our results for the rutile altermagnets can then be straightforwardly generalized for these systems.

\section{Summary}
\label{sec:summary}
Our work demonstrates that in pure altermagnets the symmetry constraint, which forces the AHE to vanish, is lifted by magnetic DWs, thereby activating a DW AHE and the orbital magnetization. In the presence of many parallel DWs, while a random arrangement of the DW chirality leads to a trivial DW AHE, we argue using Landau theory how an external magnetic field selects a chirality population that allows the AHE contribution from each DW to add up instead of canceling each other off. Our work also uncovers a crucial role of the orbital magnetization, in that it is comparable to or, in some cases, even larger than the spin magnetization. In addition, its orientation correctly indicates the Hall plane where the DW AHE is the greatest, whereas the spin magnetization fails to do so. Our work calls for experimental investigations to confirm such an important role of altermagetic DWs in Hall transport and orbital magnetism phenomena.
Due to the important role of DWs in pure altermagnets demonstrated here, future Hall-transport studies should undertake a rigorous analysis of the Hall data, ideally in conjunction with the information on the precise magnetic structure, in order to unambiguously assign an observed Hall effect to the bulk magnetic domains or to the DWs. Our work also motivates future studies of DW effects in mixed altermagnets, where bulk magnetic domains are already AHE-active and are accompanied by a weak ferromagnetism. It is probable that, in a field-sweep Hall measurement, the Hall contribution from bulk magnetic domains cancel each other off near the coercive magnetic field, exposing the DW contribution---a scenario proposed in the ferromagnet CeAlSi \cite{Piva2023}.
Future studies are also needed to firmly establish the dominant effect of the orbital magnetism uncovered here in altermagnets. This last point is further reinforced by a similar observation of a dominant orbital magnetism in MnTe in a recent first-principle study~\cite{Carmine2025}.

\section{Acknowledgement}
We thank Markus Garst, Daegeun Jo and Dongwook Go for very helpful discussions. S.S. is supported by the Deutsche Forschungsgemeinschaft (DFG) through TRR 288 Grant No. 422213477 (Project No. A11). Y.M. acknowledges support by the EIC Pathfinder OPEN grant 101129641 ``OBELIX'' and DFG TRR 173/3—268565370 (Project A11), TRR 288/2—422213477 (Project B06).

\appendix
\section{Symmetry protection of Weyl nodal lines along $\Gamma$Z and MA}
\label{sec:symmetry_protection}
In this Appendix, we discuss the symmetry protection of the Weyl nodal lines along $\Gamma$Z and MA lines seen in the band structure in Fig.\ref{fig:fig2}(b) for the case of a uniform profile $\vec N = \hat y$. As will be shown below, the Weyl nodal line along MA is protected by a combination of a nonsymmorphic mirror $g_1 = \{\mathcal{M}_x|\frac{1}{2}\frac{1}{2}\frac{1}{2}\}$ and an antiunitary symmetry $\tilde{g}_2 =\Theta \{C_{2y}|\frac{1}{2}\frac{1}{2}\frac{1}{2}\}$ which involves the time reversal $\Theta$. (Tilde on $\tilde{g}_2$ denotes its anti-unitarity.) Meanwhile, the nodal line along $\Gamma$Z is not protected by any symmetry and can be gapped out by a symmetry-conforming pertubation.

\begin{figure}[t]
	\centering
	\includegraphics[width=0.95\linewidth]{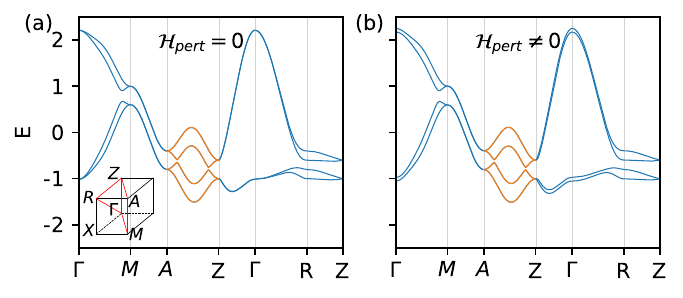}
	\caption{Demonstration how the symmetry-conforming $\ch_{\rm pert}$ gaps out the nodal line along $\Gamma$Z.}
	\label{fig:pert_band}
\end{figure}

To see the absence of a symmetry protection for the nodal line along $\Gamma$Z, we supplement the following symmetry-conforming perturbation, which takes the form of a magnetic field along x-axis, into the model Hamiltonian $\ch$ of Eq.\eqref{eq:model}
\bea
\mathcal{H}_{\rm pert} &=& \eta \sigma_x \tau_0,
\label{eq:pert}
\eea
where $\eta$ is a coefficient. One can see that $\ch_{\rm pert.}$ is symmetry-allowed from our discussion on the Landau theory for rutile altermagnets, where a N\'eel vector along y-axis induces a weak ferromagnetism with the magnetization along x-axis. Figure \ref{fig:pert_band}(b) shows how introducing $\ch_{\rm pert}$ gaps out the nodal line along $\Gamma$Z in Fig.\ref{fig:pert_band}(a). In contrast, the Weyl nodal line along MA is still protected.

Below, we outline the reason how $g_1$ and $\tilde{g}_2$ protect the MA nodal line. One can check that for all k-point along MA, i.e. $\vec k = (\pi, \pi, k_z)$, $g_1$ and $\tilde{g}_2$ are indeed the symmety. Let $\psi_{\vec k}$ be an eigenstate and $U_1$ is the unitary representation of $g_1$, then
\bea
\ch(\vec k) \psi_{\vec k} &=& E_{\vec k} \psi_{\vec k},\\
\left[\ch(\vec k), U_1\right] &=& 0,\\
U_1 \psi_{\vec k} &=& \zeta \psi_{\vec k},
\eea
where $\zeta$ is an eigenvalue of $U_1$. The square of $g_1$ is given by
\bea
g_1^2 &=& \{e|011\}\{C_{1x}|000\}.
\label{eq:squareg1}
\eea
We denote $e$ as identity and $C_{1x}$ as a 2$\pi$ rotation. This means that $\zeta$ depends on $\vec k$ and satisfies the following condition
\bea
\zeta_{\vec k}^2 &=& - e^{i\pi + ik_z},
\eea
where $-1$ arises from the 2$\pi$ rotation of the spin-$1/2$ electron, while the phase factor comes from the pure translation part of Eq.\eqref{eq:squareg1}. This means that $\zeta_{\vec k}$ takes two values
\bea
\zeta_{\vec k} &=& \pm e^{ik_z/2}. 
\eea
Next, we consider the relation between $g_1$ and $\tilde{g}_2$:
\bea
g_1 \tilde{g}_2 &=& \{e|001\}\{C_{1x}|000\}\tilde{g}_2 g_1.
\label{eq:R1}
\eea
Let $\tilde{U}_2$ be the anti-unitary representation for $\tilde{g}_2$ and, without loss of generality, let $\zeta_{\vec k}$ be $e^{ik_z/2}$, then acting $\tilde{U}_2$ on the following first equation yields:
\bea
U_1 \psi_{\vec k} &=& e^{ik_z/2} \psi_{\vec k}, \nonumber\\
\label{eq:R2}
\tilde{U}_2 U_1 \psi_{\vec k} &=& e^{-ik_z/2} \tilde{U}_2\psi_{\vec k}.
\eea
We will show that $\tilde{U}_2\psi_{\vec{k}}$ is also an eigenvector of $U_1$ but with the eigenvalue of $-e^{ik_z/2}$. To see that, we use Eq.\eqref{eq:R1}:
\bea
U_1\tilde{U}_2\psi_{\vec k} &=& -e^{ik_z} \tilde{U}_2U_1 \psi_{\vec k} = - e^{ik_z/2} \tilde{U}_2\psi_{\vec k}, 
\eea
where in the last equality, we use Eq.\eqref{eq:R2}. This means that $\tilde{U}_2 \psi_{\vec k}$ is another eigenstate of $\ch(\vec k)$ and carries the opposite $g_1$-eigenvalue to that of $\psi_{\vec k}$. Hence, it is guaranteed that there is a two-fold degeneracy protected by $g_1$ and $\tilde{g}_2$.

\begin{figure}[t]
    \centering
    \includegraphics[width=0.8\linewidth]{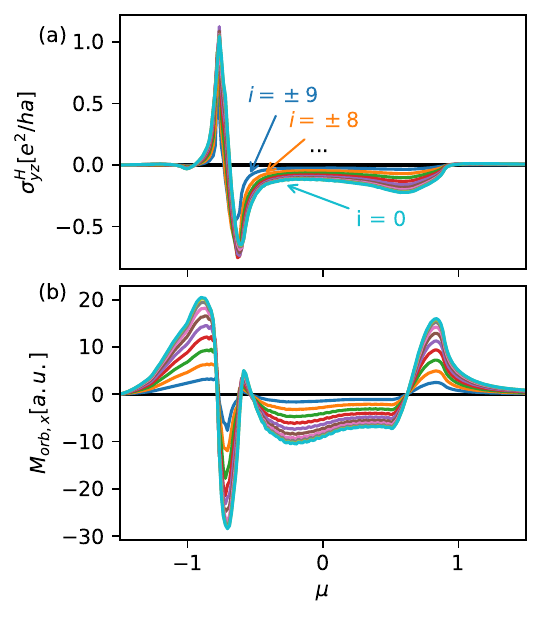}
    \caption{$\mu$-dependence of (a) $\sigma^H_{yz}(\vec N_i)$ and (b) $M_{\text{orb}, x}(\vec N_i)$ for the collinear bulk.}
    \label{fig:allcollinearbulk}
\end{figure}

\section{Kubo formula for the domain wall Hall conductivity}
\label{sec:Kubo}
The Kubo formula for the Hall conductivity is given by \cite{Sorn2021}
\begin{widetext}
\bea
    \sigma_{ij}^{H, DW} &=& \lim_{\omega \rightarrow 0} \frac{2\pi i}{L_x + 1} \frac{e^2}{h} \sum_{m, n} \int \frac{d^2 q}{(2\pi)^2} \frac{f(E_{\vec q m}) - f(E_{\vec q n})}{E_{\vec q n} - E_{\vec q m}} \left[\frac{(v_i)^{mn}_{\vec q} (v_j)^{nm}_{\vec q} }{\hbar \omega + i\gamma + E_{\vec q m} - E_{\vec q n} }\right],
    \label{eq:HallConductivity}
\eea
\end{widetext}
where $\vec q = \vec k_{\parallel} = (k_y, k_z)^T$, $\gamma$ is a small broadening, $\ket{u_{\vec q n}}$ denotes the Bloch eigenstates, and $v_j$ is the velocity operator whose matrix elements are defined below.

To obtain the velocity operator, we follow the standard procedure of the Peierls substitution: each hopping term $t_{ij} \psi^{\dagger}_i \psi_j$ acquires a phase due to the vector potential $\vec A$ associated with an external electric field $\vec E = -\partial \vec A / \partial t$:
\bea
t_{ij}\psi^{\dagger}_i\psi_j &\rightarrow & t_{ij}\psi^{\dagger}_i\psi_j \exp\left(-i\frac{e}{\hbar}\int_{\vec r_i}^{\vec r_j} d\vec r\cdot \vec A\right),\nonumber \\
&\approx& t_{ij} \psi^{\dagger}_i\psi_j \left[1 - i \frac{e}{\hbar} \vec r_{ij} \cdot \frac{\vec A(\vec r_i) + \vec A(\vec r_j)}{2}\right],
\eea
where $\vec r_{ij} = \vec r_j - \vec r_i$. The velocity operator at site $i$ is given by
\bea
    v_l(\vec r_i) &=& -\frac{1}{e}\frac{\delta H[\vec A]}{\delta A_l(\vec r_i)},
\eea
where $H[\vec A]$ is the Hamiltonian after the Peierls substitution. Finally, the velocity operator in Eq.\eqref{eq:HallConductivity} is given by
\bea
    v_l &=& \sum_i v_l(\vec r_i).
    \label{eq:tot_vel}
\eea

In the rest of the Appendix, we will apply our expressions to the case of the slab geometry and sketch the derivation of a useful form of $v_l$ Eq.\eqref{eq:tot_vel} for the Kubo formula. We will first derive the tight-binding Hamiltonian in the Fourier space and then express $v_l$ in relation to the matrix elements of the Hamiltonian.

The tight-binding Hamiltonian on the slab geometry has the following form
\bea
    H &=& \sum_{ii'll'\mu\mu'} t_{il\mu, i'l'\mu'} \psi^{\dagger}_{il\mu}\psi_{i'l'\mu'} + \text{ hc.},
\eea
where site indices $i$ and $i'$ specify the \emph{in-plane} position. The layer indices $l$ and $l'$ specify the \emph{out-of-plane} position. $\mu$ and $\mu'$ are compact indices that specify the sublattices and the spin. We note that the position vector consists of out-of-plane and in-plane components: $\vec r = (r_{\perp}, \vec{\mathfrak{r}})$.
The in-plane translational invariance is encoded in the hopping integral $t_{il\mu,i'l'\mu'} = t_{l\mu, l'\mu'}(\vec{\mathfrak{r}}_{i'} - \vec{\mathfrak{r}}_i)$, i.e. it depends only on the displacement vector $\vec{\mathfrak{r}}_{i'} - \vec{\mathfrak{r}}_{i}$ within the slab plane. The Fourier form of $H$ is given by
\bea
    H &=& \sum_{\vec q, ll'\mu\mu'} \psi^{\dagger}_{\vec q l \mu} \mathcal{H}_{l\mu, l'\mu'}(\vec q)\psi_{\vec q l'\mu'},
\eea
where
\bea
    \mathcal{H}_{l\mu, l'\mu'}(\vec q) &=& t_{l\mu, l'\mu'}(\vec q) + t_{l'\mu', l\mu}^*(\vec q)\\ 
    t_{l\mu, l'\mu'}(\vec q) &=&\sum_{i'} t_{l\mu, l'\mu'}(\vec{\mathfrak{r}}_{ii'})e^{i\vec q \cdot \vec{\mathfrak{r}}_{ii'}},
\eea
where $\vec{\mathfrak{r}}_{ii'} = \vec{\mathfrak{r}}_{i'} - \vec{\mathfrak{r}}_i$. We have also used the Fourier transformation
\bea
    \psi_{il\mu} &=& \frac{1}{\sqrt{A}} \sum_{\vec q} \psi_{\vec q l \mu} e^{i\vec q \cdot \vec{\mathfrak{r}}_i}\nonumber.
\eea
$A$ is the area of the slab.
The velocity operator Eq.\eqref{eq:tot_vel} can be computed similarly. Following Eq.\eqref{eq:tot_vel}, we can write
\bea 
    \vec{v} &=& \frac{1}{\hbar}\sum_{ii'll'\mu\mu'} it_{il\mu, i'l'\mu'} (r_{\perp, ll'} \hat x + \vec{\mathfrak{r}}_{ii'}) \psi_{il\mu}^{\dagger} \psi_{i'l'\mu'} + \text{hc.}.\nonumber \\
    && 
\eea
Taking the same Fourier transformation akin to that for the Hamiltonian, we arrive at the following form
\bea
    v_a &=& \sum_{\vec q, ll'\mu\mu'} \psi^{\dagger}_{\vec q l \mu} \mathcal{V}^a_{l\mu, l'\mu'}(\vec q) \psi_{\vec q l' \mu'}.
\eea
For the in-plane components $a = y, z,$
\bea
    \mathcal{V}^a_{l\mu, l'\mu'}(\vec q) &=& \frac{1}{\hbar}\frac{\partial \mathcal{H}_{l\mu, l'\mu}(\vec q)}{\partial q_a},
\eea
whereas the out-of-plane component $a = x$ corresponds to
\bea
    \mathcal{V}^x_{l\mu, l'\mu'}(\vec q) &=& \frac{i}{\hbar} r_{\perp, ll'} \mathcal{H}_{l\mu, l'\mu'}(\vec q).
\eea
Note that there is no summation over the indices $l$ and $l'$ in the right-hand side of the above equation.
Finally, the notation for the matrix elements of the velocity operator in Eq.\eqref{eq:tot_vel} is given by
\bea
    (v_a)^{mn}_{\vec q} &=& \bra{u_{\vec q m}} \mathcal{V}^a(\vec q) \ket{u_{\vec q n}}.
\eea

\section{Hall conductivity and orbital magnetization for collinear bulk}
\label{sec:collinearbulk}
To compute the site-averaged Hall conductivity of Eq.\eqref{eq:site_avg} and a similar site-averaged orbital magnetization, we first compute the same quantities for the collinear bulk with the N\'eel vector orientation $\vec N_i$'s that appear in the DW Ansatz for $w_{\rm dw} = 20$. Recall that $\vec N_i = \left(0, \sin \theta_i, \cos \theta_i\right)^T$ and $\theta_i = \pi(2i + w_{\rm dw})/2w_{\rm dw}$ in the DW region $-w_{\rm dw}/2 < i < w_{\rm dw}/2$. Figure \ref{fig:allcollinearbulk} shows the $\mu$-dependence of the Hall conductivity $\sigma^H_{yz}(\vec N_i)$ and that of the orbital magnetization $M_{\text{orb}, x}(\vec N_i)$. Note that $\vec N_{i = 0} = \hat y$.

\section{Non-altermagnetic origin of spin magnetization}
\label{sec:nonaltermagnetic}
In this Appendix, we illustrate how the z-component of the spin magnetization does not require altermagnetism. This can be seen by observing that the magnetic moments on the two sublattices reside at different positions, i.e. within each unit cell the sublattices A and B are located at different positions. As a result, a coarse graining of a Bloch DW profile of the N\'eel vector can qualitatively explain the non-vanishing spin magnetization. This occurs regardless of the altermagnetism or spin-orbit coupling.

\begin{figure}[t]
    \centering
    \includegraphics[width=0.9\linewidth]{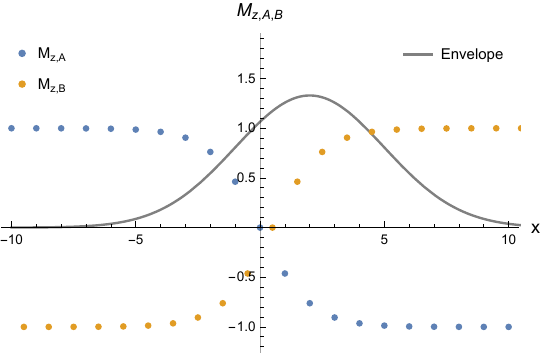}
    \caption{Profile of $M_{z,A}$ and $M_{z, B}$ near the domain wall. The solid curve illustrates an envelope function for the coarse graining.}
    \label{fig:coarsegrain}
\end{figure}

\begin{figure}[t]
    \centering
    \includegraphics[width=0.9\linewidth]{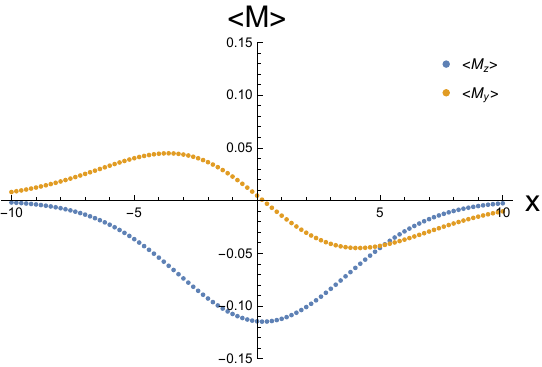}
    \caption{Profiles of $\langle M_z\rangle$ and $\langle M_y\rangle$ in a DW region. Upon summing over the domain wall region, the z-component is nontrivial, while the y-component vanishes.}
    \label{fig:nonrel_mag}
\end{figure}

\begin{figure}[t]
    \centering
    \includegraphics[width=0.8\linewidth]{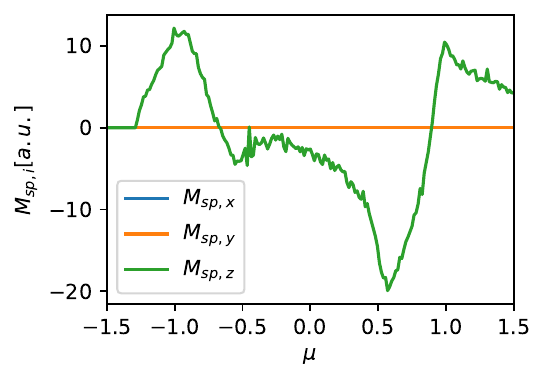}
    \caption{Spin magnetization of conduction electrons in the absence of altermagnetism and SOC, i.e. when $t_d = \lambda = \lambda' = 0$. The presence of the z-component of $\vec M_{\rm sp}$ indeed does not rely on the altermagnetism or SOC, which is in agreement with the coarse graining picture in Fig.\ref{fig:nonrel_mag}. }
    \label{fig:altermagnetism_off}
\end{figure}

We start by considering a Bloch domain wall of $\vec N$ as studied in the main text. Figure \ref{fig:coarsegrain} shows the profiles of the associated magnetic dipole moments of the sublattices A and B. Without loss of generality, we only consider their $z$-components. We then perform a coarse graining procedure to obtain an averaged local magnetization
\bea
\langle\vec M\rangle(x) &=& \sum_i \sum_{s = A, B} f(x-x_i)\vec M_s(x_i),
\eea
where $i$ is the unit-cell label, and $f(x-x_i)$ is an envelope function peaking at around $x - x_i = 0$. The solid curve in Fig. \ref{fig:coarsegrain} illustrates a Gaussian envelope function. We note that $\langle\vec M\rangle$ should be viewed as a proxy for the spin magnetization studied in the main text.

Performing the coarse graining, we obtain the profile for the $z$-component as shown in Fig.\ref{fig:nonrel_mag}. We can also repeat the same calculation for the $y$-component. Upon summing over the DW region, only does the $z$-component survives. This means that the DW profile of the N\'eel vector generally produces a nontrivial magnetization, and it relies on neither altermagnetism nor spin-orbit coupling; the same outcome arises in a conventional antiferromagnet. To illustrate the last point, we compute the spin magnetization when we switch off the altermagnetism and the spin-orbit coupling, i.e. by setting $t_d = \lambda = \lambda' = 0.$ Figure \ref{fig:altermagnetism_off} shows the $\mu$ dependence of $\vec M_{\rm sp}$ and is to be compared with Fig.\ref{fig:dwdata}(d) in the main text. The presence of z-component of $\vec M_{\rm sp}$ indeed does not rely on altermagnetism or spin-orbit coupling, and its value is qualitatively similar to that in Fig.\ref{fig:dwdata}(d).

\bibliographystyle{unsrtnat}
\bibliography{refs}

\end{document}